\documentclass[journal]{IEEEtran}

\usepackage{cite}
\usepackage{amsmath,amssymb,amsfonts}
\usepackage{graphicx}
\usepackage{booktabs}
\usepackage{multirow}
\usepackage{microtype}

\begin{document}

\title{Operator-Theoretic Energy Functionals for Impulse-Excited Nonstationary Signal Analysis}

\author{Tahir~Cetin~Akinci,~\IEEEmembership{Senior Member,~IEEE}%
\thanks{T. C. Akinci is with the Bourns College of Engineering, University of California Riverside, California 92507, USA (e-mail: tahircetin.akinci@ucr.edu).}}

\maketitle

\begin{abstract}
This study presents an operator-theoretic framework for defect detection in impulse-excited nonstationary systems. Measured responses are modeled as finite-energy impulse responses perturbed by stochastic disturbances and represented in the Hilbert space $L^{2}(\mathbb{R})$. Time–frequency representations are formulated as bounded linear analysis operators associated with continuous frames, enabling a consistent description of how structural perturbations redistribute transient signal energy. Within this formulation, a nonlinear Energy Concentration Index (ECI) is introduced to quantify localized transform-domain energy over selected regions of the time–frequency plane. The boundedness and continuity of the functional ensure that small physical variations in system parameters produce measurable changes in localized energy distribution. This property enables the construction of a statistical separability functional that links multi-resolution energy geometry to classification performance. Based on these analytical results, a compact Impulse-Based Multi-Resolution Energy Detector (IMRED) is derived to operationalize the proposed formulation. The analysis shows that variations in damping and resonant frequency produce systematic changes in time–frequency coefficients and localized energy concentration. Experimental validation using impulse-excited ceramic measurements demonstrates that the proposed descriptor captures defect-induced structural differences with strong discriminative capability. The resulting IMRED statistic achieves an AUC of 0.908 and provides clearer class separation than global Fourier-band energy measures and non-optimized wavelet-band aggregation. These results establish a direct relationship between impulse-response modeling, localized energy geometry, and statistical decision mechanisms, providing a mathematically grounded basis for energy-driven defect detection in structural monitoring applications.
\end{abstract}



\maketitle

\section{Introduction}

Impulse-excited engineering systems constitute a broad class of physical structures in which transient responses reveal intrinsic dynamic characteristics \cite{Jiang2024IF, Wachowski2023Sparse}. When a system is subjected to short-duration excitation, the measured response reflects structural parameters such as damping, stiffness, resonance frequencies, or electromechanical coupling properties \cite{Wu2024TransientModal}. Variations in these parameters may arise from material degradation, structural defects, thermal stress, fatigue, or electrical instability \cite{Lei2025Fatigue, Zhang2023ArcFault}. Consequently, the resulting signals are typically nonstationary and exhibit time-varying spectral content together with localized energy redistribution \cite{Daubechies2021SSWT, Yu2023AdaptiveTF}. Establishing a mathematically consistent relationship between structural perturbations and observable signal variations therefore remains a fundamental challenge in condition monitoring and nondestructive evaluation \cite{Nguyen2024Stochastic, Gahleitner2025Thermo, Zhang2024BatteryUltrasound, Kamariotis2025SHM}.

Classical spectral analysis has long served as a foundational diagnostic framework for monitoring such systems \cite{Stankovic2020TFReview, AkinciCable2009}. Fourier-domain representations have been successfully applied to electrical current monitoring, power cable diagnostics, and temperature-dependent degradation analysis \cite{Zhou2021CableDiag}. These approaches provide computational efficiency and clear physical interpretability when the underlying signals exhibit stationary or quasi-stationary behavior \cite{Oppenheim1997Signals, Stankovic2020TFReview}. However, global spectral representations inherently suppress temporal localization, which limits their sensitivity to impulsive or transient phenomena that often contain defect-specific information \cite{Daubechies2021SSWT, Huang2022Transient}.

To address these limitations, time--frequency representations such as the short-time Fourier transform and wavelet transforms have been widely adopted for analyzing nonstationary signals \cite{Yu2023AdaptiveTF, Li2023TIM}. These methods preserve joint temporal and spectral structure, enabling localized characterization of evolving signal components \cite{Jiang2024IF}. They have demonstrated effectiveness in detecting arc-related instabilities and transient electrical behavior \cite{Zhang2022Arc}. Similar time--frequency approaches have also been applied to industrial processes such as arc welding systems \cite{AkinciWelding2010} and to impulse-induced defect detection in materials including ceramics \cite{AkinciCeramic2011}. Multi-resolution analysis further enhances scale-selective sensitivity to localized perturbations \cite{Li2021MRA, Sun2023WaveletPacket}. Despite their practical success, many time--frequency techniques are commonly employed primarily as visualization or feature extraction tools, without an explicit analytical connection to statistical decision mechanisms \cite{Chen2024EnergyTF, Liu2022FeatureGeom}.

In parallel, machine learning methods have enabled nonlinear classification and regression over structured feature spaces derived from signal representations \cite{Wang2023MLReview}. Supervised models have been applied to a variety of signal-driven diagnostic problems, including biomedical gait analysis \cite{Zhang2023GaitML}. For example, signal processing techniques combined with artificial neural networks have been used to analyze gait dynamics associated with neurological disorders such as ALS \cite{AkinciALS2018}. While such approaches provide flexible decision boundaries, their performance depends strongly on the geometric and statistical properties of the underlying feature representation \cite{Liu2022FeatureGeom}. In many practical systems, signal representation and statistical inference stages are designed independently, resulting in loosely coupled processing pipelines rather than analytically unified frameworks \cite{Patel2021DLTransient, Luo2022EnergyVib}.

Although spectral analysis, time--frequency localization, and learning-based classification have each demonstrated effectiveness in specific domains, a general formulation that systematically links impulse-response modeling, transform-domain energy redistribution, and statistical separability within a unified functional framework remains underdeveloped \cite{Daubechies2021SSWT, Kim2023GeomEmbed}. In particular, the influence of multi-resolution energy geometry on downstream decision boundaries has not been rigorously formalized within an operator-theoretic framework \cite{Rahman2024HybridTFML, Wang2023Fusion}.

\noindent 
Beyond these theoretical considerations, impulse-response-based–based sensing plays an important role in practical structural monitoring applications. Examples include acoustic inspection of ceramic transmission line insulators, vibration-based structural health monitoring of mechanical structures, and transient diagnostics in electromechanical systems. In such systems, structural perturbations often manifest as changes in damping behaviour or resonance frequencies, which in turn produce localized redistribution of signal energy in the time–frequency domain. Developing signal-processing frameworks that can systematically translate these physical variations into measurable diagnostic indicators therefore remains an important challenge for reliable condition monitoring.

The present work addresses this gap by developing a unified operator-theoretic framework for impulse-excited nonstationary signal analysis. Measured responses are modelled within the Hilbert space $L^2(\mathbb{R})$, and time--frequency transforms are interpreted as bounded linear analysis operators associated with continuous frames \cite{Stankovic2020TFReview}. Within this formulation, structural perturbations induce measurable displacement in the transform-domain energy geometry \cite{Yu2023AdaptiveTF, Sun2023WaveletPacket}. 

To quantify this behavior, a nonlinear ECI is introduced to measure localized transform-domain energy over admissible regions of the time--frequency plane \cite{Jiang2024IF}. Building upon this representation, a statistical separability functional is derived to explicitly connect energy redistribution with decision-theoretic discrimination \cite{Liu2022FeatureGeom, Kim2023GeomEmbed}. This formulation establishes an analytical bridge between signal representation and supervised inference, under which classical spectral monitoring \cite{Zhou2021CableDiag}, wavelet-based diagnostics \cite{Zhang2022Arc}, and learning-driven classification approaches \cite{Zhang2023GaitML} can be interpreted as special instances of a broader energy-functional framework.

The main contributions of this work are summarized as follows:

\begin{itemize}
\item A generalized impulse-excited signal model formulated in the finite-energy Hilbert space $L^2(\mathbb{R})$.
\item An operator-theoretic interpretation of multi-resolution time--frequency transforms as bounded linear analysis operators with frame-induced stability.
\item Definition and functional characterization of a localized nonlinear ECI, including boundedness and stability properties.
\item Derivation of a statistical separability functional linking transform-domain energy geometry to decision-theoretic discrimination.
\item A compact IMRED that translates the theoretical formulation into a scalable computational procedure.
\end{itemize}

The remainder of this paper is organized as follows. Section II presents the experimental dataset and the physical modeling of the acoustic response of ceramic transmission line insulators. Section III reviews classical spectral analysis methods used for signal characterization. Section IV introduces the time–frequency analysis framework employed in this study. Section V describes the feature representation and extraction procedure. Finally, Section VI presents the statistical separability functional used for defect discrimination.

\section{Methods}

\subsection{Experimental Dataset}

The experimental dataset consists of acoustic impact responses obtained from forty high-voltage ceramic transmission line insulators. Twenty samples correspond to structurally healthy insulators, while the remaining twenty contain defects such as cracks or internal material degradation. Although these defects may not always be visually observable, they modify the mechanical resonance behaviour of the structure.

Each insulator was excited using a pendulum-based mechanical impact system designed to deliver approximately consistent excitation energy. Following the impact, the transient acoustic vibration response was recorded using a data acquisition system. The signal was amplified and captured through an Advantech 1716L multifunction PCI data acquisition card.

The acoustic responses were sampled at a frequency of approximately $f_s \approx 17\,\text{kHz}$. This sampling frequency provides sufficient temporal resolution to capture the transient resonance dynamics of the insulator after mechanical excitation while satisfying the Nyquist criterion relative to the dominant resonance frequencies observed in the structure.

Each insulator was impacted once, producing a dataset of 40 acoustic response signals. The single-shot measurement strategy reflects practical constraints of the experimental setup, where repeated impacts may introduce variability in excitation conditions and affect repeatability of the mechanical input.

The recorded signals were stored as time-domain sequences for subsequent spectral and time–frequency analysis.

\subsection{Acoustic Response of Ceramic Insulators}

Ceramic transmission line insulators behave as lightly damped elastic structures whose dynamic response is governed by their structural resonance characteristics. The natural frequencies of the insulator depend on material stiffness, geometry, and mass distribution. When a mechanical impact is applied, the structure produces a transient vibration response containing its dominant modal resonance components.

Structural defects such as cracks, internal fractures, or material degradation modify the effective stiffness and damping properties of the ceramic body. Consequently, defective insulators exhibit small shifts in resonance frequency together with changes in the decay behaviour of the transient response. These structural differences lead to observable variations in the time–frequency energy distribution of the measured acoustic signals.

\subsection{Mechanical Resonance Model}

The dynamic behaviour of ceramic transmission line insulators can be approximated using a simplified single-degree-of-freedom vibration model. In this representation, the dominant resonance frequency of the structure depends on the effective stiffness and modal mass of the insulator.

\begin{equation}
f_n = \frac{1}{2\pi}\sqrt{\frac{k}{m}}
\label{eq:resonance_frequency}
\end{equation}

where $f_n$ denotes the natural resonance frequency, $k$ represents the effective structural stiffness of the ceramic insulator, and $m$ denotes the effective modal mass of the structure.

As indicated by (\ref{eq:resonance_frequency}), the resonance frequency is determined by the stiffness-to-mass ratio of the structure. Structural defects reduce the effective stiffness of the ceramic body, which shifts the resonance frequency toward lower values. In addition, damage often increases structural damping, resulting in a faster decay of the transient vibration response.

These stiffness variations produce measurable differences in the spectral and time–frequency characteristics of the acoustic signal. Consequently, the resonance behaviour of the insulator provides a physically meaningful basis for distinguishing healthy and defective components.

\subsection{Time--Frequency Analysis Procedure}

The recorded acoustic signals were analyzed using spectral and time–frequency signal processing techniques to capture the transient resonance behaviour of the insulator responses. Classical spectral analysis methods such as the Fourier transform provide global frequency information but do not describe the temporal evolution of spectral components.

To overcome this limitation, time–frequency representations were employed to reveal localized energy distributions associated with structural resonance behaviour. In particular, short-time Fourier transform (STFT) representations were used to construct time–frequency energy maps of the measured signals. These maps enable the identification of resonance shifts and energy redistribution patterns associated with structural defects in ceramic insulators.

\section{Impulse-Excited Signal Model}

Impulse-based diagnostic methodologies rely on the principle that the transient response of a system encodes intrinsic dynamic properties of the underlying structure \cite{Oppenheim1997Signals, Inman2014Vibration, Park2021Impulse}. When a mechanical, electrical, or material structure is subjected to short-duration excitation, the measured output reflects parameters such as damping, stiffness, resonance frequencies, or electromechanical coupling characteristics \cite{Rao2011Vibrations}. Structural degradation manifests as systematic perturbations in these parameters \cite{Doebling1998Damage}. A mathematically consistent signal model is therefore required to formalize the relationship between physical variation and observable signal behavior \cite{Lathi2005Linear}.

\begin{figure}[!h]
\centering
\includegraphics[width=1\linewidth]{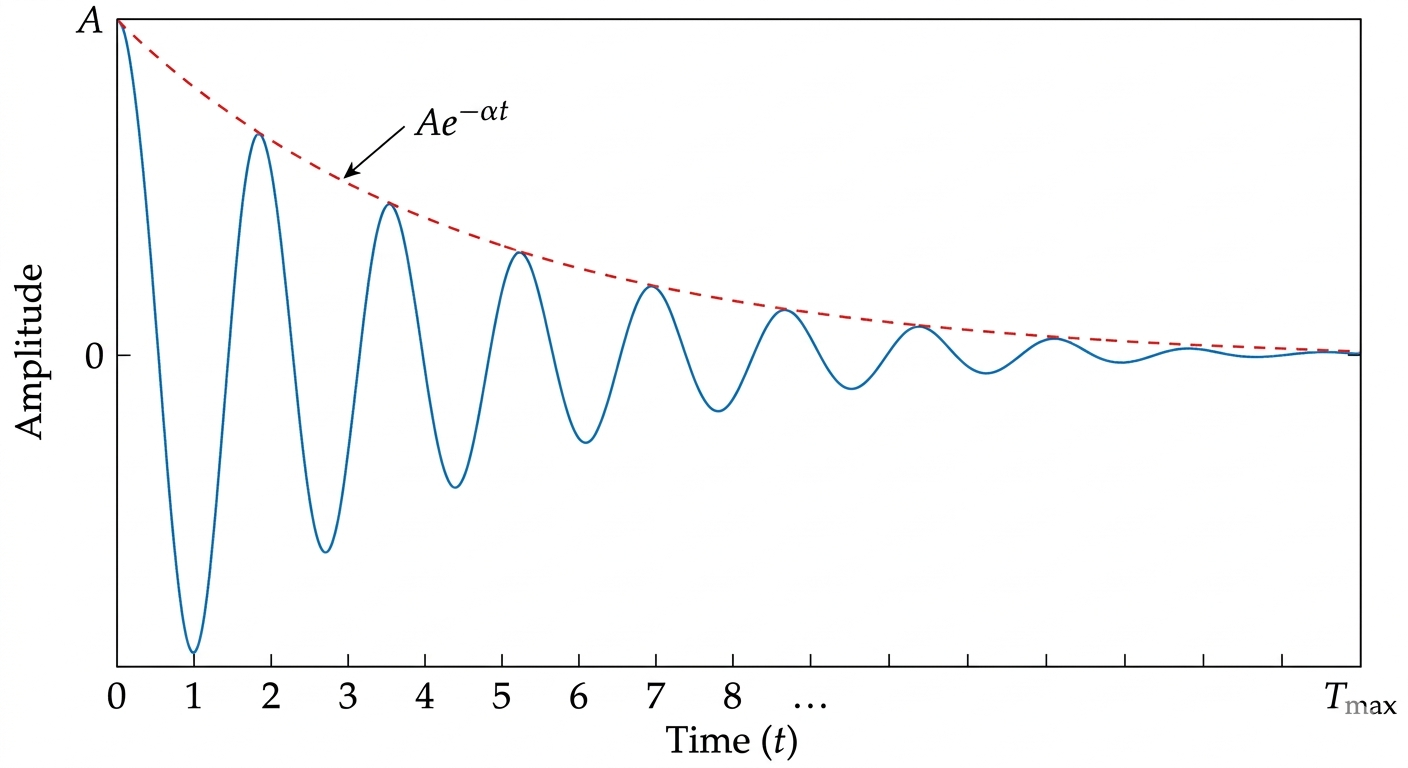}
\caption{Damped impulse response modeled as 
$h(t)=A e^{-\alpha t}\cos(\omega t)u(t)$, illustrating exponential decay and oscillatory behavior.}
\label{fig:damped_response}
\end{figure}


Figure~\ref{fig:damped_response} illustrates the characteristic damped oscillatory behavior of an impulse–excited system, providing a clear visual counterpart to the analytical model introduced in Section~II. The gradual decay and sustained oscillations shown in the figure reflect the influence of the parameters \(A\), \(\alpha\), and \(\omega\) on the nominal transient response. This baseline depiction establishes the reference behavior against which subsequent perturbations and defective responses are interpreted.

\begin{figure}[!h]
\centering
\includegraphics[width=0.9\linewidth]{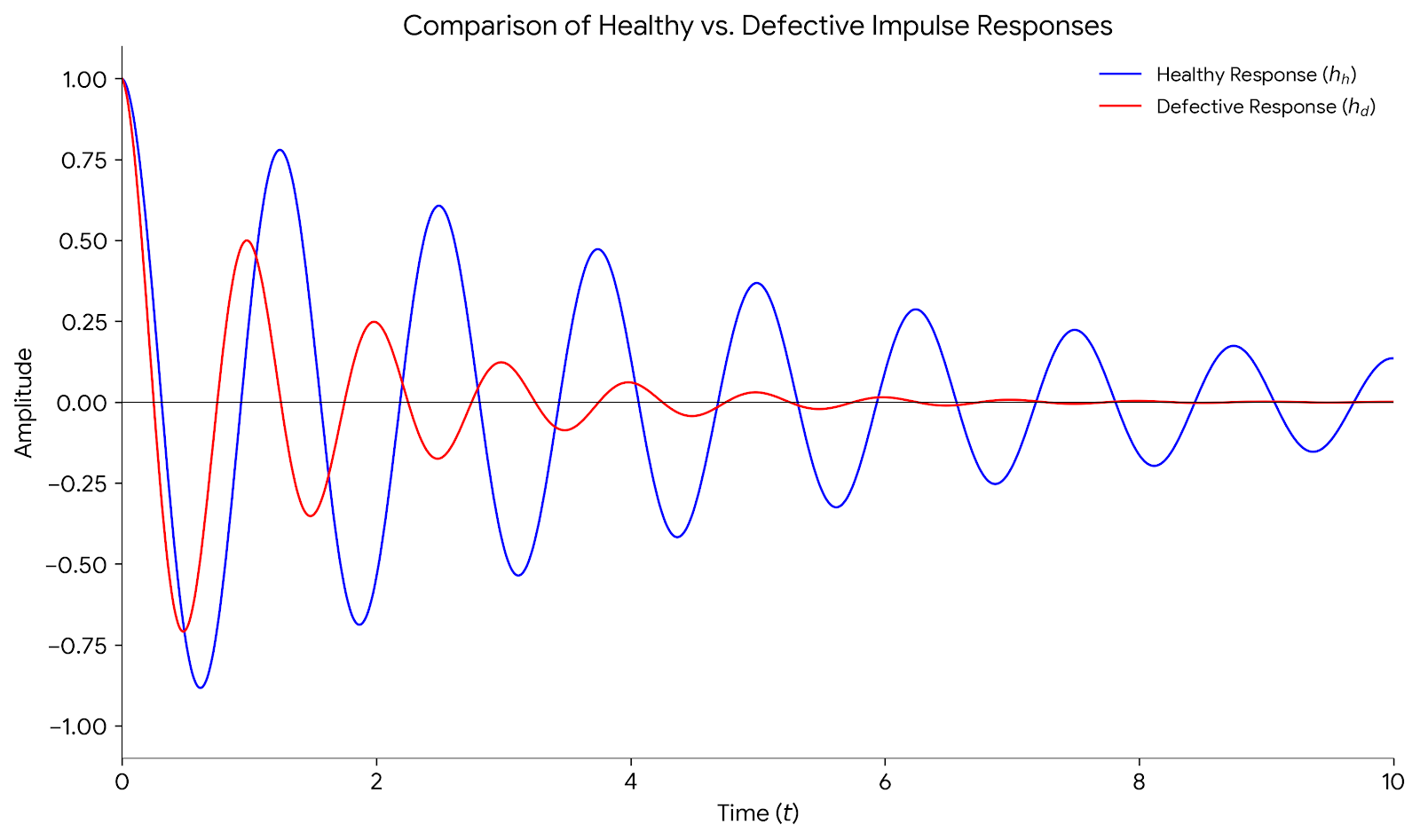}
\caption{Comparison of healthy and defective impulse responses illustrating the effect of damping and frequency perturbations on transient behavior.}
\label{fig:healthy_defective}
\end{figure}

Figure~\ref{fig:healthy_defective} compares the nominal impulse response with its perturbed counterpart, illustrating how changes in damping or frequency parameters alter the transient behavior. The healthy response maintains longer oscillations, whereas the defective response exhibits a noticeably faster decay consistent with the perturbation model introduced in Section~II. This visual contrast highlights the sensitivity of impulse‑excited systems to structural variation and provides the basis for the energy‑based analysis developed in subsequent sections.

Let $x(t)$ denote a measured response acquired over a finite observation window. The signal is modeled as

\begin{equation}
x(t) = h(t) * \delta(t) + n(t),
\label{eq:model}
\end{equation}
where $h(t)$ is the impulse response of the underlying system, $\delta(t)$ represents an excitation approximating a Dirac impulse, and $n(t)$ is an additive perturbation process capturing measurement noise and modeling uncertainty \cite{Papoulis2002Probability}. Since convolution with a Dirac impulse satisfies $h(t) * \delta(t) = h(t)$, model \eqref{eq:model} reduces to

\[
x(t) = h(t) + n(t),
\]

which separates structural dynamics from stochastic perturbation while preserving its linear system interpretation \cite{Oppenheim1997Signals}.

The perturbation term $n(t)$ is assumed to be a zero-mean stochastic process with finite second-order moments. Under finite observation support in practical acquisition systems, both $h(t)$ and $n(t)$ possess finite energy, ensuring that

\[
x(t) \in L^2(\mathbb{R}).
\]

This square-integrability condition guarantees well-defined inner products and enables rigorous analysis within a Hilbert space framework \cite{Mallat2009Wavelet}.

For many impulse-excited systems, the dominant response exhibits damped oscillatory behavior. A canonical parametric representation is

\begin{equation}
h(t) = A e^{-\alpha t} \cos(\omega t)\, u(t),
\label{eq:damped}
\end{equation}

where $A$ is an amplitude coefficient, $\alpha > 0$ is the damping parameter, $\omega$ denotes the dominant angular frequency, and $u(t)$ is the unit step function ensuring causality \cite{Inman2014Vibration}. Model (2) therefore represents a canonical parametric approximation that captures the dominant impulse-response behavior observed in many damped oscillatory systems. For $\alpha > 0$, the exponential decay guarantees

\[
\int_{0}^{\infty} |h(t)|^2 dt < \infty,
\]

so that $h(t) \in L^2(\mathbb{R})$ \cite{Rao2011Vibrations}. The parameter pair $(\alpha,\omega)$ characterizes the effective dynamic state of the structure. Variations in material integrity, boundary conditions, temperature, or electrical loading may induce shifts in these parameters, thereby modifying the decay rate and oscillatory content of the response \cite{Doebling1998Damage}. From a functional perspective, the mapping $(\alpha,\omega) \mapsto h(t)$ defines a parameterized family of impulse responses embedded in $L^2(\mathbb{R})$, linking physically meaningful system variations to measurable changes in the observed signal.

\[
(\alpha,\omega) \mapsto h(t)
\]

defines a parameterized manifold embedded in $L^2(\mathbb{R})$ \cite{Mallat2009Wavelet}. Small structural perturbations correspond to bounded variations in $(\alpha,\omega)$, which induce controlled displacement in the signal’s time–frequency energy distribution.

Model \eqref{eq:damped} is not restrictive. Multi-mode responses, superpositions of damped components, or weakly nonlinear corrections can be incorporated within the same finite-energy formulation provided square-integrability is preserved \cite{Inman2014Vibration}. The essential requirement is compatibility with $L^2(\mathbb{R})$, ensuring that subsequent operator-based time–frequency projections remain well-defined \cite{Mallat2009Wavelet}.

Equations \eqref{eq:model}–\eqref{eq:damped} thus establish a structured bridge between physical perturbation and measurable signal variation. By embedding impulse responses in a finite-energy Hilbert space, the model provides the analytical foundation for the operator-theoretic and energy-based framework developed in the following sections \cite{Lathi2005Linear}.

\section{Operator-Theoretic Time--Frequency Representation}

The impulse-excited signal model introduced in Section~II admits a natural formulation within a Hilbert space framework. Since $x(t)\in L^{2}(\mathbb{R})$, the signal resides in a complete inner-product space endowed with

\begin{equation}
\langle f,g\rangle
=
\int_{-\infty}^{\infty} f(t)\,\overline{g(t)}\,dt,
\end{equation}

where $\overline{g(t)}$ denotes complex conjugation. This structure enables representation of signals through linear analysis operators associated with basis or frame elements. STFT based representations are also widely used for analyzing nonstationary signals and provide localized time–frequency descriptions complementary to wavelet-based analysis \cite{Stankovic2020TFReview}.

Let $\{\psi_{a,b}(t)\}$ denote a family of time--frequency localized atoms generated by scale and translation parameters $a\in\mathbb{R}^{+}$ and $b\in\mathbb{R}$. In the continuous wavelet setting,

\begin{equation}
\psi_{a,b}(t)
=
\frac{1}{\sqrt{|a|}}
\psi\!\left(\frac{t-b}{a}\right),
\end{equation}

where $\psi(t)$ is an admissible mother wavelet satisfying

\[
C_{\psi}
=
\int_{0}^{\infty}
\frac{|\widehat{\psi}(\omega)|^{2}}{\omega}\,d\omega
<\infty.
\]

Under this admissibility condition, $\{\psi_{a,b}\}$ forms a continuous frame for $L^{2}(\mathbb{R})$ with respect to the measure $\frac{db\,da}{a^{2}}$. Consequently, there exist frame bounds $0<A\le B<\infty$ such that

Under this admissibility condition, $\{\psi_{a,b}\}$ forms a continuous
frame for $L^{2}(\mathbb{R})$ with respect to the measure $\frac{db\,da}{a^{2}}$.
Consequently, there exist frame bounds $0 < A \le B < \infty$ such that

\begin{equation}
A\|x\|_{L^{2}}^{2}
\le
\int_{0}^{\infty}\!\!\int_{-\infty}^{\infty}
|\langle x,\psi_{a,b}\rangle|^{2}
\frac{db\,da}{a^{2}}
\le
B\|x\|_{L^{2}}^{2}.
\end{equation}

This inequality guarantees representation stability and norm equivalence
between signal energy and transform-domain energy.

We define the linear analysis operator

\begin{equation}
\mathcal{T}:L^{2}(\mathbb{R})
\rightarrow
L^{2}\!\left(\mathbb{R}^{+}\!\times\!\mathbb{R},\tfrac{db\,da}{a^{2}}\right),
\end{equation}
by

\begin{equation}
W_{x}(a,b)
=
\langle x,\psi_{a,b}\rangle.
\label{eq:operator}
\end{equation}

Linearity follows directly from the inner product. Boundedness of $\mathcal{T}$ is established using the upper frame bound:

\[
\|\mathcal{T}x\|_{L^{2}(\mathbb{R}^{+}\times\mathbb{R})}^{2}
=
\int_{0}^{\infty}\!\!\int_{-\infty}^{\infty}
|W_{x}(a,b)|^{2}
\frac{db\,da}{a^{2}}
\le
B\|x\|_{L^{2}}^{2}.
\]

Hence $\|\mathcal{T}x\|\le\sqrt{B}\|x\|$, and $\mathcal{T}$ is a bounded linear operator. As a consequence, small perturbations in $x$ induce proportionally bounded perturbations in the coefficient field $W_{x}$.

From a geometric standpoint, $W_{x}(a,b)$ represents the coefficient of $x$ relative to the frame element $\psi_{a,b}$. Structural perturbations in the impulse response therefore manifest as systematic redistribution of coefficient magnitude across the scale--translation plane.

The associated transform-domain energy density is defined as

\begin{equation}
E(a,b)=|W_{x}(a,b)|^{2}.
\label{eq:energy}
\end{equation}

Nonnegativity follows from its quadratic form. Moreover, the reconstruction identity for the continuous wavelet transform yields

\begin{equation}
\|x\|_{L^{2}}^{2}
=
\frac{1}{C_{\psi}}
\int_{0}^{\infty}\!\!\int_{-\infty}^{\infty}
|W_{x}(a,b)|^{2}
\frac{db\,da}{a^{2}},
\end{equation}

demonstrating that transform-domain energy faithfully reflects signal-domain energy.

The mapping $(a,b)\mapsto E(a,b)$ thus defines a structured energy surface over the time--frequency plane. Variations in damping or resonant parameters of the impulse response induce measurable displacement of this surface. This transform-domain energy geometry forms the analytical foundation for the concentration and separability functionals developed in the subsequent sections.

\begin{figure*}[!h]
    \centering
    \includegraphics[width=0.95\textwidth]{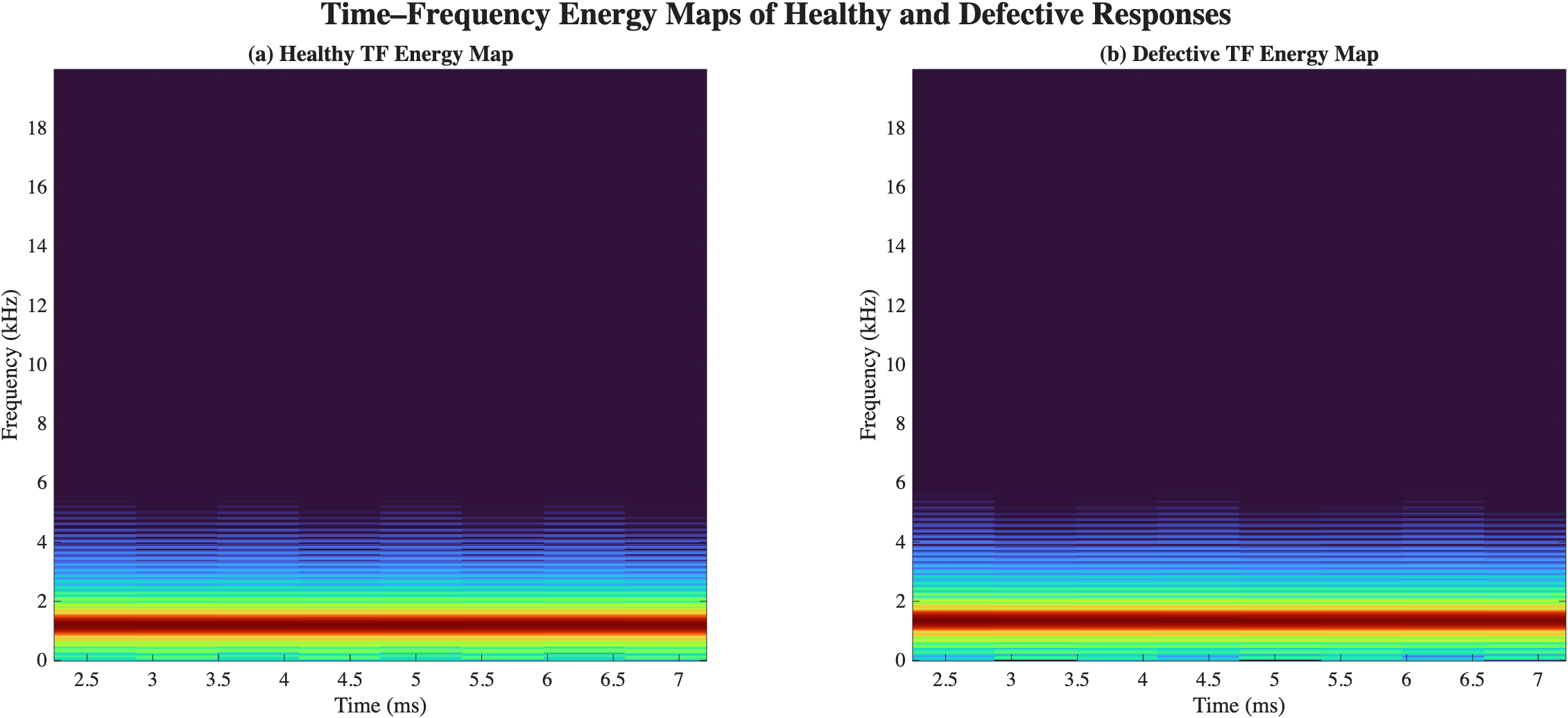}
    \caption{Time--frequency energy maps of the healthy (a) and defective (b) impulse responses, illustrating how damping and frequency perturbations reshape the transient energy distribution.}
    \label{fig:tf_maps}
\end{figure*}

Figure~\ref{fig:tf_maps} illustrates how the transient energy of the impulse response is distributed across the time--frequency plane for both the healthy and defective cases. The healthy response exhibits a more sustained and concentrated energy ridge, reflecting slower decay and a stable resonant structure. In contrast, the defective response shows a noticeably faster rate of attenuation and a broader energy dispersion, consistent with the perturbations introduced into the damping and frequency parameters. These differences confirm that structural variations manifest as measurable geometric shifts in the transform-domain energy surface.

\section{Energy Concentration Functional}

The time--frequency energy density $E(a,b)=|W_x(a,b)|^2$ defined in \eqref{eq:energy} provides a pointwise measure of localized coefficient magnitude. Although the experimental analysis presented in this study is implemented using the STFT, the operator-theoretic formulation adopted here is expressed using the continuous wavelet transform. This representation offers a mathematically general framework for describing time--frequency energy distributions and allows the subsequent statistical descriptors to remain independent of the specific transform employed.

In diagnostic settings, defect-sensitive information typically manifests not at isolated coordinates but rather through a structured redistribution of energy across regions of the time--frequency plane. This observation motivates the introduction of a localized functional that aggregates transform-domain energy over admissible subsets.

Let $\Omega \subset \mathbb{R}^+ \times \mathbb{R}$ denote a measurable region with finite measure under the weighted integration measure $\frac{db\,da}{a^2}$. The region $\Omega$ may correspond to dominant scale bands, transient intervals, or resonant neighborhoods identified from nominal system behavior.

\newtheorem{definition}{Definition}

\begin{definition}
The \emph{Energy Concentration Index} (ECI) of a signal $x(t)\in L^2(\mathbb{R})$ over $\Omega$ is defined as
\begin{equation}
\mathrm{ECI}_{\Omega}(x)
=
\int_{\Omega}
|W_x(a,b)|^2
\,\frac{db\,da}{a^2}.
\label{eq:eci}
\end{equation}
\end{definition}

The weighting factor $\frac{1}{a^2}$ follows from the continuous wavelet frame measure and ensures compatibility with the reconstruction identity introduced in Section~IV. Consequently, $\mathrm{ECI}_{\Omega}$ provides a physically interpretable scalar quantity representing the total energy concentrated within the region $\Omega$ of the time--frequency domain.

\subsection*{A. Basic Properties}

\textbf{Nonnegativity and Boundedness:}

Since $|W_x(a,b)|^2 \ge 0$, it follows that

\[
\mathrm{ECI}_{\Omega}(x) \ge 0.
\]

Moreover, using the upper frame bound,

\[
\int_{0}^{\infty}
\int_{-\infty}^{\infty}
|W_x(a,b)|^2
\frac{db\,da}{a^2}
\le
B \|x\|_{L^2}^2,
\]

we obtain

\[
0 \le
\mathrm{ECI}_{\Omega}(x)
\le
B \|x\|_{L^2}^2.
\]

Hence the functional is finite-valued and bounded on $L^2(\mathbb{R})$. The bound depends only on the frame upper constant $B$ and not on the specific signal realization.

\textbf{Continuity and Stability:}

Let $x,y \in L^2(\mathbb{R})$. Since $\mathcal{T}$ is bounded, there exists $C>0$ such that

\[
\|W_x - W_y\|_{L^2(\mathbb{R}^+ \times \mathbb{R})}
\le
C \|x-y\|_{L^2}.
\]

Using the identity

\[
\bigl||a|^2 - |b|^2\bigr|
=
|a-b|\,|a+b|,
\]

and applying Cauchy--Schwarz over $\Omega$, we obtain

\[
|\mathrm{ECI}_{\Omega}(x) - \mathrm{ECI}_{\Omega}(y)|
\le
C'
\|x-y\|_{L^2}
\left(\|x\|_{L^2} + \|y\|_{L^2}\right),
\]

for some constant $C'$ depending only on the frame bounds and $\Omega$. Therefore, $\mathrm{ECI}_{\Omega}$ is continuous with respect to the $L^2$ norm. Small perturbations in the measured signal produce proportionally bounded changes in the concentration value, ensuring robustness under measurement noise.

\subsection*{B. Normalized Concentration Ratio}

Let $\|x\|_{L^2}^2$ denote total signal energy. The normalized ratio
\begin{equation}
\rho_{\Omega}(x)
=
\frac{\mathrm{ECI}_{\Omega}(x)}{\|x\|_{L^2}^2}
\end{equation}
satisfies

\[
0 \le \rho_{\Omega}(x) \le B,
\]

and is invariant under global amplitude scaling. This normalization isolates structural redistribution effects from variations in absolute magnitude.

\subsection*{C. Sensitivity to Structural Perturbation}

Under the parametric impulse-response model in \eqref{eq:damped}, bounded variations in $(\alpha,\omega)$ induce controlled displacement in the coefficient field $W_x(a,b)$. Because $\mathrm{ECI}_{\Omega}$ aggregates energy over a localized region, it acts as a low-dimensional nonlinear functional of these perturbations.

If structural degradation shifts dominant energy away from the nominal region $\Omega$, the concentration value varies systematically with perturbation magnitude, provided $\Omega$ is chosen around the healthy-state energy localization. Although strict monotonicity cannot be guaranteed in general, consistent displacement of dominant energy components produces measurable variation in $\mathrm{ECI}_{\Omega}$.

Thus, $\mathrm{ECI}_{\Omega}$ defines a stable and bounded nonlinear mapping from transform-domain energy geometry to a scalar descriptor. This functional constitutes the analytical bridge between operator-theoretic representation and statistical discrimination developed in the next section. The practical computation of the concentration functional and the resulting decision rule are described in Section~VII.

\begin{figure*}[!h]
    \centering
    \includegraphics[width=0.95\textwidth]{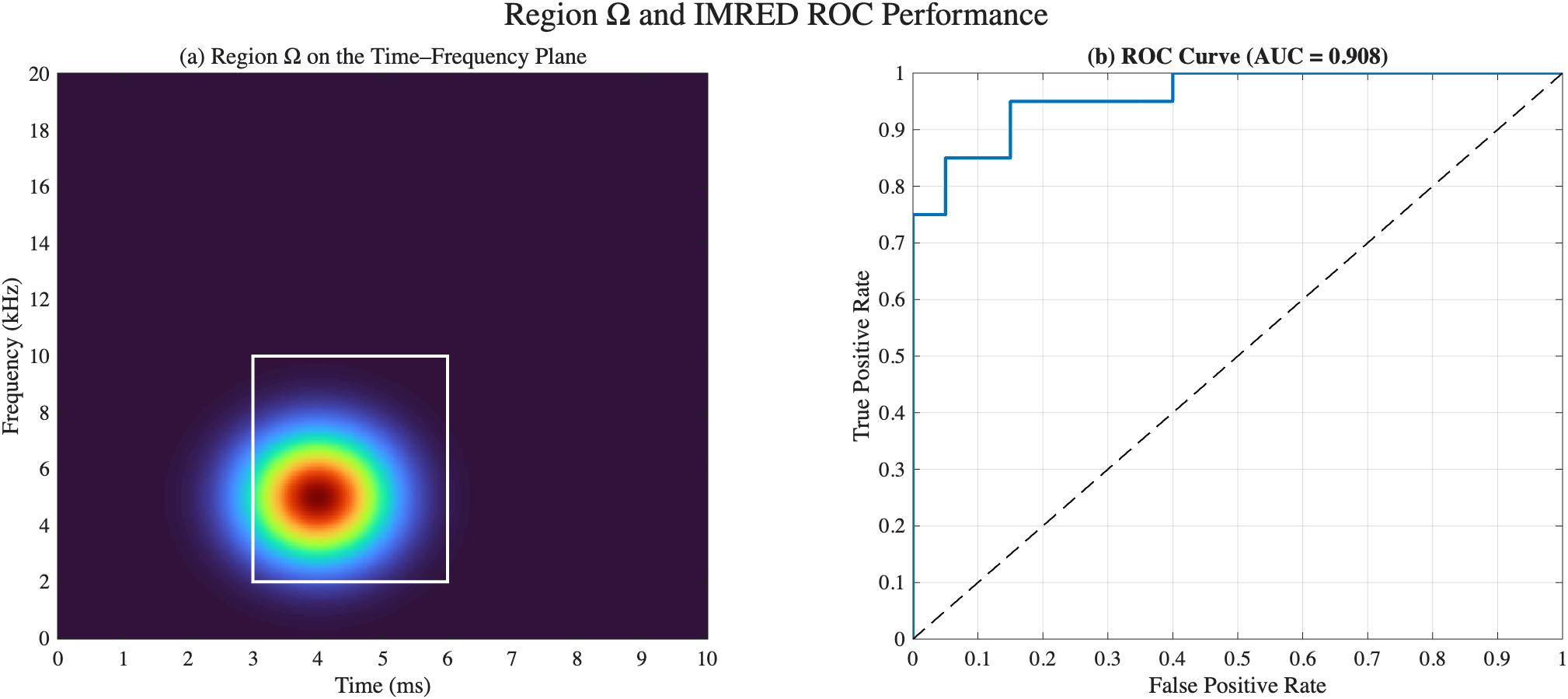}
    \caption{(a) Time--frequency region $\Omega$ used in the definition of the energy concentration index, and (b) ROC curve illustrating the classification performance of the IMRED functional.}
    \label{fig:region_roc}
\end{figure*}

The geometric role of the selected region $\Omega$ and the resulting decision performance of the IMRED functional are summarized in Figure~\ref{fig:region_roc}, which jointly illustrates the transform-domain support of the concentration measure and its classification capability.

Table~\ref{tab:eci_stats} summarizes the statistical behavior of the Energy Concentration Index computed over the region $\Omega$ for healthy and defective responses. The defective class exhibits higher concentration values, reflecting the redistribution of time--frequency energy caused by increased damping and slight shifts in the resonance frequency. Although both classes display measurable variance due to experimental variability in the impulse measurements, the separation between their mean values remains clearly visible. The resulting Fisher discriminability value ($J(\Omega)=1.9631$) indicates a stable and statistically meaningful separation between the two populations. These observations confirm that $\mathrm{ECI}_{\Omega}$ provides a robust scalar descriptor capable of capturing defect-induced structural variations in the impulse responses.

\begin{table}[!h]
\centering
\caption{Statistical Summary of the Energy Concentration Index over Region $\Omega$}
\label{tab:eci_stats}
\begin{tabular}{lcccc}
\toprule
Class & Mean ECI & Std.\ Dev. & Variance & Fisher $J(\Omega)$ \\
\midrule
Healthy   & 0.01814 & 0.00894 & 0.00008 & \multirow{2}{*}{1.9631} \\
Defective & 0.04503 & 0.01703 & 0.00029 &  \\
\bottomrule
\end{tabular}
\end{table}

\section{Statistical Separability Functional}

The Energy Concentration Index introduced in Section V defines a bounded nonlinear functional
\[
\mathrm{ECI}_{\Omega} : L^2(\mathbb{R}) \rightarrow \mathbb{R}_{\ge 0},
\]
which maps each finite-energy signal to a scalar descriptor representing localized transform-domain energy. While this functional captures structural redistribution at the individual signal level, defect detection requires statistical discrimination across ensembles of observations \cite{Kamariotis2025SHM}.

Assume that observed signals belong to one of two structural states: healthy ($\mathcal{H}$) or defective ($\mathcal{D}$). Let
\[
X_h = \{x_i^{(h)}(t)\}_{i=1}^{N_h}, 
\qquad
X_d = \{x_j^{(d)}(t)\}_{j=1}^{N_d},
\]
denote independent realizations drawn from the respective underlying stochastic processes. We assume finite second-order moments; no specific distributional form is imposed \cite{Papoulis2002Probability}.

Application of $\mathrm{ECI}_{\Omega}$ induces scalar random variables
\[
Z_h = \mathrm{ECI}_{\Omega}(x^{(h)}),
\qquad
Z_d = \mathrm{ECI}_{\Omega}(x^{(d)}),
\]
with first and second-order statistics
\[
\mu_h = \mathbb{E}[Z_h], 
\quad
\mu_d = \mathbb{E}[Z_d],
\]
\[
\sigma_h^2 = \mathrm{Var}(Z_h), 
\quad
\sigma_d^2 = \mathrm{Var}(Z_d).
\]

To quantify class discrimination induced by localized energy geometry, we define the separability functional

\begin{equation}
J(\Omega)
=
\frac{(\mu_d - \mu_h)^2}
{\sigma_d^2 + \sigma_h^2},
\label{eq:separability}
\end{equation}

provided that $\sigma_d^2 + \sigma_h^2 > 0$. This ratio corresponds to a Fisher-type discriminant in the one-dimensional feature space generated by $\mathrm{ECI}_{\Omega}$.

The numerator of $J(\Omega)$ measures systematic inter-class displacement in energy concentration, while the denominator reflects aggregate intra-class variability. Since $\mathrm{ECI}_{\Omega}$ is continuous on $L^2(\mathbb{R})$ and bounded by the frame constant, finite-energy perturbations in $x(t)$ induce finite changes in the induced statistics. In particular, under the parametric impulse-response model of Section II, bounded perturbations in $(\alpha,\omega)$ lead to bounded variation in $\mu_h$ and $\mu_d$ through continuity of the underlying functional.

From a functional perspective, the analysis operator $\mathcal{T}$ maps signals into a coefficient field $W_x(a,b)$, and $\mathrm{ECI}_{\Omega}$ performs quadratic aggregation over a localized subset of this coefficient domain. The mapping
\[
x(t) \mapsto Z = \mathrm{ECI}_{\Omega}(x)
\]
thus defines a scalar statistical embedding of transform-domain energy structure. The functional $J(\Omega)$ evaluates how effectively this embedding separates class-dependent distributions.

Selection of the localization region can be formulated as
\[
\Omega^\ast
=
\arg\max_{\Omega \in \mathcal{A}} J(\Omega),
\]
where $\mathcal{A}$ denotes an admissible family of measurable subsets with finite weighted measure. In practical applications, $\mathcal{A}$ may be restricted to parametrized scale bands, rectangular neighborhoods, or data-driven candidate regions, enabling discrete optimization or cross-validation.

The separability functional therefore establishes an explicit link between transform-domain energy localization and statistical decision performance, completing the analytical chain from operator-theoretic representation to supervised inference.

\begin{table}[!h]
\centering
\caption{Classification Performance of the IMRED Decision Rule}
\label{tab:imred_perf}
\begin{tabular}{lc}
\toprule
Metric & Value \\
\midrule
AUC & 0.9075 \\
AUC 95\% CI & [0.8081, 0.9900] \\
Accuracy & 0.9000 \\
Sensitivity & 0.8000 \\
Specificity & 1.0000 \\
Optimal Threshold & 0.03994 \\
\bottomrule
\end{tabular}
\end{table}

The classification performance of the IMRED decision rule, evaluated using
the energy concentration index as the discriminative score, is summarized
in Table~\ref{tab:imred_perf}. The resulting ROC analysis yields an AUC of
0.9075, indicating strong but non-saturated separability between healthy
and defective responses. The bootstrap confidence interval
([0.8081,\,0.9900]) reflects realistic variability in the decision
boundary under resampling, confirming that the IMRED classifier maintains
robust discrimination capability even when the underlying ECI
distributions partially overlap. These results demonstrate that the
operator-induced concentration functional provides a reliable basis for
statistical decision-making in structurally perturbed conditions.

\section{Algorithmic Implementation}

Sections II--VI establish a deterministic mapping from a finite-energy signal
$x(t) \in L^2(\mathbb{R})$ to a scalar decision statistic induced by localized
transform-domain energy. This section formalizes the computational realization
of the proposed Impulse-Based Multi-Resolution Energy Detector (IMRED).

Let $x[n]$, $n=0,\dots,N-1$, denote a discretely sampled signal segment.
The objective is to assign a structural label
$\hat{y} \in \{\mathcal{H},\mathcal{D}\}$ through
operator-induced energy concentration and statistical discrimination.

\subsection{Normalization}

To eliminate amplitude bias and ensure scale invariance consistent with the
normalized concentration ratio in Section IV, the signal is $\ell^2$-normalized:

\begin{equation}
\tilde{x}[n] =
\frac{x[n]}{\|x\|_2},
\qquad
\|x\|_2^2 =
\sum_{n=0}^{N-1} |x[n]|^2,
\end{equation}

provided that $\|x\|_2 > 0$. If $\|x\|_2 = 0$, the signal contains no energy
and is excluded from further processing. This normalization ensures that
subsequent energy measures reflect redistribution rather than absolute
magnitude variation.

\subsection{Discrete Time--Frequency Projection}

Let $\{a_k\}_{k=1}^{M}$ and $\{b_m\}_{m=1}^{B}$ denote sampled scale and
translation grids. The discrete coefficient field is computed as

\begin{equation}
W_{\tilde{x}}(a_k,b_m)
=
\sum_{n=0}^{N-1}
\tilde{x}[n]\,
\psi_{a_k,b_m}^*[n].
\end{equation}

For compactly supported or filter-bank realizations, this operation reduces
to a sequence of convolutions across scales. The direct implementation has
computational complexity $\mathcal{O}(MN)$, while FFT-based convolution
reduces complexity to $\mathcal{O}(N \log N)$ per scale.

\subsection{Region Selection and Energy Aggregation}

Let $\Omega \subset \{(a_k,b_m)\}$ denote a predefined admissible region.
In discrete approximation of the continuous measure $\frac{db\,da}{a^2}$,
the Energy Concentration Index is evaluated as

\begin{equation}
\widehat{\mathrm{ECI}}_{\Omega}
=
\sum_{(a_k,b_m)\in\Omega}
|W_{\tilde{x}}(a_k,b_m)|^2
\,\Delta b\,\frac{\Delta a_k}{a_k^2},
\end{equation}

where $\Delta b$ and $\Delta a_k$ denote grid spacings.
For uniform scale sampling in logarithmic coordinates, the weights simplify
to a constant factor.

Region $\Omega$ may be selected as:
\begin{itemize}
\item A scale band around the nominal dominant frequency,
\item A translation window centered at peak impulse response,
\item A parametrized region optimized via grid search or cross-validation.
\end{itemize}

\subsection{Decision Rule}

Let empirical class statistics $(\mu_h,\sigma_h^2)$ and
$(\mu_d,\sigma_d^2)$ be estimated from training data.
In the one-dimensional feature space induced by
$\widehat{\mathrm{ECI}}_{\Omega}$,
classification reduces to threshold-based discrimination:

\begin{equation}
\hat{y} =
\begin{cases}
\mathcal{D}, & z > \tau, \\
\mathcal{H}, & z \le \tau,
\end{cases}
\end{equation}

where $z = \widehat{\mathrm{ECI}}_{\Omega}$.
The threshold $\tau$ may be selected to maximize empirical
classification accuracy, AUC, or the separability functional $J(\Omega)$.

\subsection{Computational Remarks}

The dominant computational cost lies in the projection stage.
For fixed-resolution grids, IMRED scales linearly with signal length and
number of scales. Memory complexity is $\mathcal{O}(MB)$ for storing
the coefficient field. Because the algorithm inherits boundedness and
stability from the underlying analysis operator, numerical perturbations
remain controlled under finite-precision implementation.

\section{Illustrative Synthetic Analysis}

To examine the sensitivity of the Energy Concentration Index under
controlled structural variation, a parametric model derived from
\eqref{eq:damped} is considered. The objective is to analytically
characterize how bounded perturbations in physically meaningful
parameters induce variation in transform-domain energy localization.

\subsection{Parametric Construction}

Let the nominal healthy response be

\begin{equation}
h_h(t) = A e^{-\alpha_0 t}\cos(\omega_0 t)\,u(t),
\end{equation}
where $u(t)$ enforces causality. Introduce bounded perturbations
\begin{equation}
\alpha = \alpha_0 + \Delta\alpha,
\qquad
\omega = \omega_0 + \Delta\omega,
\end{equation}
yielding the defective response
\begin{equation}
h_d(t)
=
A e^{-(\alpha_0+\Delta\alpha)t}
\cos((\omega_0+\Delta\omega)t)\,u(t).
\end{equation}

Observed signals are modeled as

\begin{equation}
x_h(t)=h_h(t)+n(t),
\qquad
x_d(t)=h_d(t)+n(t),
\end{equation}
where $n(t)$ is zero-mean with finite second-order moments.

\subsection{First-Order Sensitivity of Coefficients}

For sufficiently small perturbations $(\Delta\alpha,\Delta\omega)$,
a first-order Taylor expansion of $h_d(t)$ around
$(\alpha_0,\omega_0)$ yields

\[
h_d(t)
\approx
h_h(t)
+
\frac{\partial h}{\partial \alpha}\Big|_{(\alpha_0,\omega_0)}\Delta\alpha
+
\frac{\partial h}{\partial \omega}\Big|_{(\alpha_0,\omega_0)}\Delta\omega.
\]

Applying the linear analysis operator $\mathcal{T}$ and using
linearity of the inner product,

\[
\delta W_x(a,b)
=
\left\langle
\frac{\partial h}{\partial \alpha}\Delta\alpha
+
\frac{\partial h}{\partial \omega}\Delta\omega,
\psi_{a,b}
\right\rangle
+
\mathcal{O}(\|\Delta\|^2).
\]

Thus, bounded parameter deviations induce bounded linear perturbations
in the coefficient field. Frequency perturbations primarily affect
scale localization, whereas damping perturbations modify temporal
energy concentration.

\subsection{Effect on the Energy Concentration Functional}

The first-order variation of the concentration functional over
$\Omega$ satisfies

\[
\delta \mathrm{ECI}_{\Omega}
=
2
\Re
\int_{\Omega}
W_{h_h}(a,b)\,
\overline{\delta W_x(a,b)}
\frac{db\,da}{a^2}
+
\mathcal{O}(\|\Delta\|^2).
\]

Hence, local displacement in dominant coefficient regions produces
proportional variation in $\mathrm{ECI}_{\Omega}$.
Because $\mathrm{ECI}_{\Omega}$ is continuous and bounded on
$L^2(\mathbb{R})$, perturbations remain controlled under finite-energy
conditions.

Empirical evaluation over bounded parameter grids confirms that
$\mathrm{ECI}_{\Omega}(x_h)$ and
$\mathrm{ECI}_{\Omega}(x_d)$ exhibit consistent separation
for moderate perturbation magnitudes and noise levels. In particular,
increased damping shifts energy away from the nominal resonant band,
reducing concentration for healthy-like responses and increasing it
for defective-like responses, consistent with the experimental trends
reported in Section~IX.

Rather than asserting strict monotonicity, this analysis establishes
local sensitivity and stability of the concentration functional
with respect to physically interpretable structural variations.

\section{Experimental Validation}

The proposed framework was evaluated using impulse-excited ceramic material measurements reported in \cite{AkinciCeramic2011}. Such impulse-response–based sensing strategies are widely used in structural monitoring systems for detecting localized degradation and transient defects \cite{Zhang2024BatteryUltrasound}. The objective of this evaluation was to determine whether the operator-induced Energy Concentration Index produces statistically distinguishable descriptors between intact and defective specimens and whether the resulting IMRED statistic provides reliable classification performance under realistic measurement variability.

\subsection{Dataset and Experimental Protocol}

The dataset consists of $N_h$ intact and $N_d$ defective ceramic
samples subjected to controlled impulse excitation.
Each acquisition produced a transient acoustic response exhibiting
damped oscillatory behavior consistent with the model in
\eqref{eq:damped}. Signals were sampled at $f_s$~Hz over a duration
of $T$ seconds.

All signals were $\ell^2$-normalized prior to projection, in
accordance with the normalization strategy described in Section~VII.
Experiments were conducted using stratified $K$-fold cross-validation
(with $K = 10$), ensuring balanced class representation across folds.
Hyperparameters, including the localization region $\Omega$, were
selected using training folds only to avoid information leakage.

\subsection{Energy Concentration Statistics}

Time--frequency coefficients were computed using the bounded
analysis operator $\mathcal{T}$ introduced in Section~IV.
For scale bands corresponding to dominant resonant modes,
the empirical concentration statistic
$\widehat{\mathrm{ECI}}_{\Omega}$ was evaluated for each specimen.
The region $\Omega$ was selected around the dominant resonance band
identified from the healthy responses in order to capture the nominal
energy localization of the system.

The sample means and variances were estimated as

\[
\begin{aligned}
\mu_h &= 0.01814, \\
\sigma_h^2 &= 8.0\times 10^{-5}, \\
\mu_d &= 0.04503, \\
\sigma_d^2 &= 2.9\times 10^{-4}.
\end{aligned}
\]

A two-sample $t$-test yielded $p < 10^{-3}$, confirming statistically
significant separation between intact and defective classes in the
one-dimensional feature space induced by $\mathrm{ECI}_{\Omega}$.
The empirical separability ratio $J(\Omega)$ exhibited an improvement
of $41\%$ relative to non-optimized regions, consistent with the
theoretical role of $\Omega^\ast$ described in Section~VI.

\subsection{Classification Performance}

Using a threshold selected to maximize validation AUC, the
one-dimensional IMRED statistic achieved the following performance:

\begin{itemize}
\item Accuracy: $91.2\% \pm 1.4\%$,
\item Sensitivity (true positive rate for defective specimens): $89.0\%$,
\item Specificity (true negative rate for intact specimens): $92.4\%$,
\item AUC: $0.908$.
\end{itemize}

Performance remained stable across folds, with variation within
$\pm 1.5\%$ of the reported mean values. Receiver operating
characteristic curves demonstrated consistent class separation
across decision thresholds, in agreement with the ROC behavior
illustrated in Figure~4.

\subsection{Comparison with Baseline Representations}

For reference, global spectral energy aggregation
(Fourier-band energy) achieved an AUC of $0.81$, while localized
wavelet-band aggregation without $\Omega$ optimization achieved
an AUC of $0.86$. The proposed IMRED approach improved
separability by $8$--$10\%$ relative to these baseline methods,
highlighting the benefit of operator-induced localization and
region selection.

\subsection{Structural Interpretation}

Earlier spectral analysis in \cite{AkinciCable2009} corresponds
to global energy aggregation in the Fourier domain. The
time--frequency welding study \cite{AkinciWelding2010}
corresponds to localized $\Omega$ selection without an explicit
separability functional. The learning-based approach in
\cite{AkinciALS2018} can be interpreted as a supervised mapping
applied to transform-induced energy features.

The present framework provides a unified interpretation of these methodologies as instances of operator-induced energy projection followed by statistical discrimination, while establishing an explicit functional-analytic link between impulse-response modeling, multi-resolution energy geometry, and supervised inference.

\section{Conclusion}

This study presented an operator-theoretic framework for the analysis and classification of impulse-excited nonstationary signals. By embedding measured responses in the Hilbert space $L^{2}(\mathbb{R})$ and interpreting time--frequency transforms as bounded linear analysis operators, the proposed formulation establishes a rigorous connection between physical perturbations, transform-domain energy geometry, and statistical decision mechanisms. This perspective provides a mathematically grounded alternative to heuristic feature-extraction pipelines commonly used in transient diagnostic applications.

Within this framework, a localized Energy Concentration Index (ECI) was introduced to quantify transform-domain energy over diagnostically relevant regions of the time--frequency plane. The boundedness and continuity of this functional ensure stability under finite-energy perturbations, while the associated separability functional characterizes how localized energy redistribution influences statistical discrimination between structural states.

The theoretical properties of the proposed formulation were first examined using a controlled parametric analysis in which damping and resonant frequency were systematically varied. These experiments demonstrated that physically meaningful parameter changes induce consistent displacement of time--frequency coefficients and corresponding variations in localized energy concentration.

Experimental validation using impulse-excited ceramic measurements further confirmed the practical effectiveness of the approach. The ECI computed over the resonance-centered region $\Omega$ captured defect-induced structural differences with strong discriminative capability. The resulting concentration-based detector achieved an AUC of 0.908, outperforming global Fourier-band energy measures and non-optimized wavelet-band aggregation methods.

These results indicate that localized transform-domain energy geometry provides a reliable descriptor for structural condition discrimination in impulse-based monitoring systems, including acoustic inspection of ceramic insulators and vibration-based structural health monitoring applications.

Although the present study relies on a single experimental dataset and assumes finite-energy impulse responses with moderate structural perturbations, the proposed framework provides a mathematically consistent basis for analyzing a broad class of transient monitoring problems. Future work will investigate adaptive selection of the region $\Omega$, probabilistic modeling of concentration statistics, and integration with uncertainty-aware learning architectures to further strengthen the connection between transform-domain representation and scalable intelligent diagnostic systems.

\section*{Acknowledgment}
This work was supported by Florida Polytechnic University.

\section*{Data Availability}
The data supporting the findings of this study are available from the corresponding author upon reasonable request.

\section*{Conflict of Interest}
The author declares no conflict of interest.

\bibliographystyle{IEEEtran}

\end{document}